\documentclass[journal,twoside,web]{ieeecolor}
\usepackage{tmi}
\usepackage{cite}
\usepackage{amsmath,amssymb,amsfonts}
\usepackage{algorithmic}
\usepackage{graphicx}
\usepackage{array}
\usepackage{textcomp}
\usepackage{multirow}
\usepackage[final]{changes}
\usepackage{fancyhdr}

\begin{document}
\title{An Annotation-free Restoration Network for Cataractous Fundus Images}
\author{Heng Li, Haofeng Liu, Yan Hu, Huazhu Fu, Yitian Zhao, Hanpei Miao, Jiang Liu
\thanks{This work was supported in part by Guangdong Provincial Department of Education (2020ZDZX3043), Guangdong Provincial Key Laboratory (2020B121201001), Guangdong Basic and Applied Fundamental Research Fund Committee (2020A1515110286), and Shenzhen Natural Science Fund (JCYJ20200109140820699 and 20200925174052004). (Corresponding Author: Yan Hu and Jiang Liu)}
\thanks{H. Li, Y. Hu, H. Liu, H. Miao are with the Department of Computer Science and Engineering, Southern University of Science and Technology, Shenzhen 518055, China (e-mail: lih3, huy3@sustech.edu.cn; 12032880, miaohp@mail.sustech.edu.cn). H. Li and H. Liu contributed equally to this manuscript.}
\thanks{Y. Zhao is with the Cixi Institute of Biomedical Engineering, Ningbo Institute of Industrial Technology, Chinese Academy of Sciences, Ningbo 315201, China (e-mail: yitian.zhao@nimte.ac.cn).}
\thanks{H. Fu is with the Institute of High Performance Computing, Agency for Science, Technology and Research, Singapore. (e-mail: hzfu@ieee.org). 
}
\thanks{J. Liu is with Research Institute of Trustworthy Autonomous Systems, Department of Computer Science and Engineering, Southern University of Science and Technology, Shenzhen 518055, China (e-mail: liuj@sustech.edu.cn).}
\thanks{Corresponding authors: Y. Hu, J. Liu.}
\thanks{Code of this study is available at https://github.com/liamheng/ Restoration-of-Cataract-Images-via-Domain-Adaptation.}
}

\maketitle


\begin{abstract}
Cataracts are the leading cause of vision loss worldwide. Restoration algorithms are developed to improve the readability of cataract fundus images in order to increase the certainty in diagnosis and treatment for cataract patients. Unfortunately, the requirement of annotation limits the application of these algorithms in clinics. This paper proposes a network to annotation-freely restore cataractous fundus images (ArcNet) so as to boost the clinical practicability of restoration. Annotations are unnecessary in ArcNet, where the high-frequency component is extracted from fundus images to replace segmentation in the preservation of retinal structures. The restoration model is learned from the synthesized images and adapted to real cataract images. Extensive experiments are implemented to verify the performance and effectiveness of ArcNet. Favorable performance is achieved using ArcNet against state-of-the-art algorithms, and the diagnosis of ocular fundus diseases in cataract patients is promoted by ArcNet. The capability of properly restoring cataractous images in the absence of annotated data promises the proposed algorithm outstanding clinical practicability.
\end{abstract}


\section{Introduction}
\label{sec:introduction}
\begin{figure}[tp]
\begin{centering}
\includegraphics[width=8.6cm]{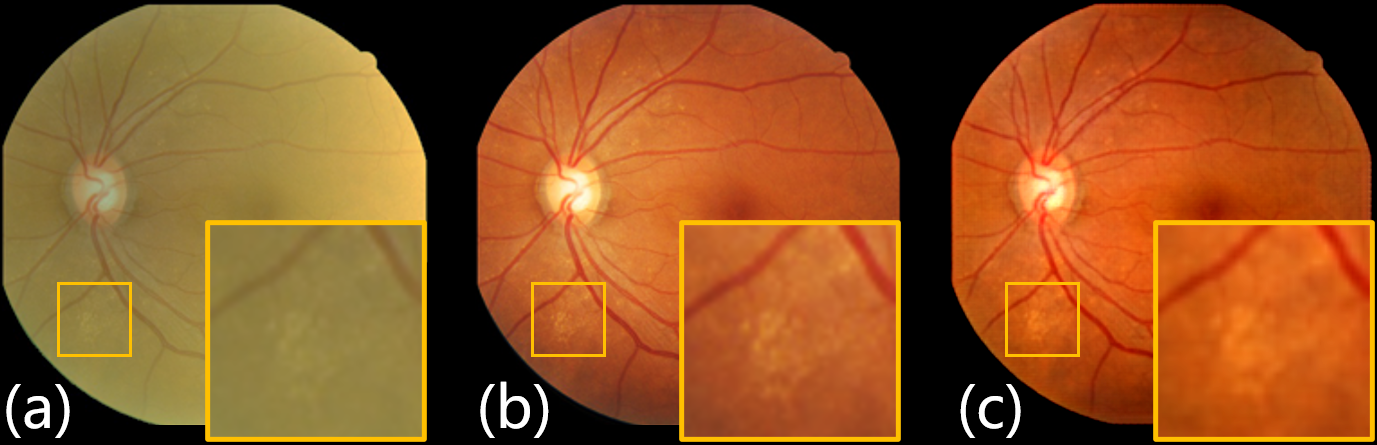}
\par\end{centering}
\caption{Illustration of restoration for the cataractous image. (a) a cataractous fundus image before surgery. (b) the corresponding image after surgery. (c) the image restored from the cataractous one by ArcNet. The restoration enhances the abnormality in yellow boxes.} \label{fig:introduction}
\end{figure}

\IEEEPARstart{C}{ataracts}, the leading cause of vision loss in the world, account for 33.4\% blindness and 18.4\% moderate-to-severe visual impairment~\cite{flaxman2017global,lee2017global}. 
Surgery is the most cost-effective and commonly used treatment of cataracts and has been used to cure more than 30 million people worldwide in 2020~\cite{world2015blindness}. 
Although the technology of cataract surgery has gradually advanced recently, treating cataract patients with ocular comorbidities is still cumbersome for the clinician.
For instance, due to the sensitivity reduction and mesopic visual function resulting from glaucoma, intraocular lenses transplanted in cataract surgery might cause significant vision disturbances in patients with glaucoma~\cite{kumar2007multifocal}.
For patients with diabetic retinopathy, cataract surgery often increases the risk of macular edema~\cite{flesner2002cataract}, which is a major cause of vision loss after cataract surgery~\cite{kim2007analysis}.
Therefore, a complete and comprehensive preoperative examination is crucial in cataract surgery to discriminate ocular pathologies that could negatively influence the postoperative outcomes~\cite{alio2017multifocal}.
However, fundus assessment of cataract patients is a challenging task for clinicians. 
Since fundus images are captured through the lens, cataracts will attenuate and scatter the light passing through the lens degrading a photograph. 
To make matters worse, unlike the degradation caused by environmental noise, the degradation from cataracts is a pathological characteristic, which is unavoidable in fundus images.
Hence, the restoration of fundus photographs taken through cataracts is clinically valuable~\cite{peli1987enhancement}. As the example shown in Fig.~\ref{fig:introduction}, cataracts haze the fundus abnormalities before surgery, compared to the fundus image taken after surgery.
And the proposed restoration algorithm enhances the observation of the abnormality in the cataractous fundus image.

To clearly visualize the fundus for reliable assessment and treatment, extensive work has long been done to enhance the cataractous fundus image~\cite{li2021applications}. According to the imaging principle of fundus photography, Peli et al.~\cite{peli1989restoration} developed an optical model for imaging the fundus through cataracts and proposed a restoration scheme for cataractous images using a homomorphic Wiener filter. 
Contrast enhancement is an efficient paradigm to promote the readability of images, by limiting contrast amplification, contrast limited adaptive histogram equalization (CLAHE) has been presented~\cite{zuiderveld1994contrast} and introduced to restore degraded fundus images~\cite{mitra2018enhancement,setiawan2013color}.
For boosting the observations of the fundus, promising denoizing algorithms, such as dark channel prior~\cite{he2010single} and guided image filtering (GIF)~\cite{he2012guided}, have been modified to restore the fundus images obtained through cataracts~\cite{cheng2018structure}.
However, due to the complexity of the imaging process of fundus photography and the mechanism of deterioration, it is challenging for these hand-crafted algorithms to preserve pathological characteristics and restore fundus images, simultaneously. 
The recent advancement in deep learning techniques has facilitated the application of convolutional neural networks (CNN) in image restoration~\cite{ren2019low,cai2016dehazenet,huang2021neighbor2neighbor}, and specific approaches have been proposed to improve the quality of degraded fundus images~\cite{zhao2019data,shen2020understanding,luo2020dehaze}. 

Although these restoration approaches have achieved remarkable performance, practical difficulties appear when introducing them to the cataractous images.
a) Collecting restoration annotations for cataractous fundus images, i.e. cataract-free images from identical patients, is costly and impractical in clinics. Therefore, the restoration algorithms demanding supervised data are unsuitable for cataractous images.
b) Structure preservation by segmentation provides sub-optimal results. 
Preserving structures by segmentation~\cite{shen2020understanding,luo2020dehaze} not only aggravates annotation costs, but even worse, leads to missing unannotated pathological characteristics in restorations.
c) These learned restoration models suffer significant performance drop encountering domain shifts. Domain shifts, referring to the changes in the data distribution between the training dataset and the dataset to deploy, are common and troublesome in clinics.

To develop a practical restoration scheme for cataractous images, an end-to-end network is proposed in this paper to learn the restoration model of clinical cataractous images without any annotations. 
Primarily,  a source domain of high-low quality paired data is synthesized by simulating cataract-like images from publicly available data for the training of the restoration network. 
Subsequently,  the segmentation annotation is replaced by the structure guidance of the high-frequency component (HFC), which is acquired from the fundus images, to preserve retinal structures during the restoration.
Then a restoration network is designed for the fundus images obtained through cataracts using image-to-image translation and unsupervised domain adaptation. 
Finally, the comparisons with state-of-the-art algorithms and the applications for promoting fundus assessment were presented, while cataractous fundus images from four datasets were employed to demonstrate the performance of the proposed network. 
The proposed approach is significantly modified from our previous work published in ISBI-2021~\cite{li2021Restoration}, where the simulation model for cataracts has been improved and the HFC has replaced the Sobel operator as the structure guidance.
The main contributions of this paper are summarized as follows:
\begin{itemize}
  \item [1)] 
  A network for annotation-free restoration of cataractous fundus images (ArcNet) is proposed to promote the retina examination for cataract patients. Considering the acquirement of any annotation is impractical in clinical scenarios, ArcNet presents a practical solution for fundus restoration.
  \item [2)]
  Instead of segmentation, the HFC is extracted from fundus images to guide the restoration, where fundus characteristics are automatically and comprehensively preserved.
  \item [3)]
  The restoration model learned from the synthesized data is properly generalized to the cataractous images by unsupervised domain adaptation.
  \item [4)]
  Extensive experiments validated the promising performance of the proposed ArcNet. Compared to the state-of-the-art restoration algorithms, ArcNet achieved favorable performance, and its effectiveness was demonstrated in quality restoration and diagnosis boosting.
\end{itemize}

\section{Related Work}
Fundus photography, a routine examination for fundus assessment, tends to experience quality degradation in the attendance of cataracts, bias illumination, or unskilled operation. 
Since Peli et al. verified the feasibility and clinical value for enhancing the fundus images~\cite{peli1987enhancement}, and proposed an optical model for imaging the fundus photography through cataracts~\cite{peli1989restoration}, restoration algorithms have been developed to analyze fundus images better.

\subsection{Fundus image enhancement}
CLAHE is an efficient technique for contrast enhancement and has been extensively applied to retinal fundus images~\cite{setiawan2013color,geetharamani2016retinal, zhou2017color}. Mitra et al.~\cite{mitra2018enhancement} combined CLAHE with Fourier transform to enhance the contrast of cataractous images.
Alternatively, to enhance retinal fundus images, efforts have been made to estimate and compensate for the noise in imaging models. 
He~\cite{he2012guided} proposed GIF to filter images as an edge-preserving smoothing operator and efficiently compute haze removed outputs.
Cheng et al.~\cite{cheng2018structure} developed a structure-preserving guided retinal image filtering (SGRIF), which modified GIF to preserve the structure in retinal images. 
Frequency domain approaches are another paradigm of image enhancement. 
Low-pass frequency filtering and $\alpha$-rooting were conducted to improve the contrast of the retinal structures~\cite{cao2020retinal}. 
Although these methods perform efficiently, their heavy dependence on global image statistics leads to severe constraints in implementation.

\subsection{CNN-based restoration algorithms}
\label{sec:CNNs}
Recently, owing to the advantage in image representation, CNN has become the most fashionable algorithm in the community of computer vision. 
Methods for promoting image quality have rapidly advanced based on CNN through image enhancement~\cite{ren2019low,lore2017llnet}, dehazing~\cite{cai2016dehazenet,chen2018robust}, and denoizing~\cite{lehtinen2018noise2noise,huang2021neighbor2neighbor}. Unfortunately, due to the differences between medical and natural images, the fundamental assumption of these schemes may not hold in fundus images.
As preserving retinal structures is significant in retinal image analysis, pixel-wise translation has thus been introduced to retinal fundus image enhancement. 
The networks applied to retinal image enhancement are composed of two categories: paired~\cite{isola2017image} and unpaired~\cite{zhu2017unpaired} translation.
As the paired ones require image pairs to learn a mapping from one representation to another, degradation was simulated to build retinal image pairs of high-low quality~\cite{perez2020conditional,shen2020understanding}. 
CycleGAN utilizing unpaired data to model the translation, is one of the most frequently-used translation networks~\cite{engin2018cycle} and has been implemented in the restoration of fundus images ~\cite{zhao2019data,yoo2020cyclegan}. Further, for preserving retinal structures, segmentation was introduced to enhance retinal structures.
A clinical-oriented fundus enhancement network (CofeNet) was proposed~\cite{shen2020understanding} to preserve the retinal structures in restoration according to segmentation outputs.
Luo et al.~\cite{luo2020dehaze} reported a two-stage dehazing algorithm, which restores cataractous fundus images under the supervision of segmentation. 
However, annotations of restoration and segmentation are necessary to implement these algorithms.
\subsection{Domain adaptation}
Most statistical learning algorithms strongly depend on the assumption that the source (train) and target (test) data are independent and identically distributed~\cite{zhou2021domain}, while ignoring the out-of-distribution (OOD) scenarios commonly in practice. Due to the domain shift between the source and target domain, a model learned from source data will suffer significant performance drops on an OOD target domain~\cite{motiian2017unified}.
Domain adaptation aims to bridge the shift between the source and target domain~\cite{tzeng2017adversarial,yang2020fda}, such that the model learned from the source domain is able to perform decently on the target data.

Domain adaptation algorithms are divided into three categories. The first one attempts to learn a mapping between source and target domains~\cite{hoffman2018cycada,zhao2019multi}. The second one tries to find a shared latent space for source and target distributions~\cite{tzeng2017adversarial,luo2019taking}. And the third one regularizes a classifier trained on a source distribution to work well on a target distribution~\cite{motiian2017unified,yu2021divergence}. Supervision is necessary for the third one, while unsupervised domain adaptation requiring no annotated target data always follows the first two categories. Recently, with the advance of deep learning techniques, adversarial learning has been extensively employed to align source and target distributions in the image~\cite{hoffman2018cycada}, output~\cite{tsai2018learning}, or latent space~\cite{tzeng2017adversarial}.

Unsupervised domain adaptation is imported to achieve an annotation-free restoration network in this study. Cataract-like images are first simulated to build a source domain close to the target one, and then a shared network is implemented with unsupervised domain adaptation to generalize the restoration model from the source to the target domain.

\section{Method}
\label{sec:method}
To mitigate the data requirement of annotations, in this study automatically captured structure guidance and unsupervised domain adaptation are introduced to restore unannotated cataractous fundus images. 

An overview of the proposed ArcNet is exhibited in Fig. \ref{fig:overview} (a). High-low quality paired data are synthesized for training the restoration model by simulating cataract-like images from any available clear ones. HFC in fundus images is extracted as structure guidance to preserve retinal structures during restoration. Image-to-image translation and unsupervised domain adaptation are cooperated to learn the restoration model from the source domain of simulated data and adapt to the target domain of real images. Thus, the unannotated cataractous images can be unsupervisedly restored.

\subsection{Cataract-like image simulation}
According to the imaging model proposed in~\cite{peli1989restoration}, the fundus image $I$ degraded by cataracts is composed of retinal reflectance information and light reflected back from the cataract: 
\begin{equation}
I(i,j)=\alpha\cdot L\cdot \gamma(i,j)\cdot\tau (i,j)+L(1-\tau(i,j)),
\label{eq:imaging}
\end{equation}
\noindent where $(i,j)$ indicates pixels in the image, $\alpha$ is the attenuation constant of retinal illumination. $L$ is the illumination of the fundus camera, $\gamma(i,j)$ and $\tau(i,j)$ respectively denote the retinal reflectance function and the transmission function of the lens. Approaches to estimate $\tau(i,j)$ were proposed in previous studies~\cite{he2012guided,cheng2018structure}, but their application scenarios are quite limited in restoration.

The source domain of high-low quality image pairs is synthesized by simulating cataract-like images, to learn the restoration model for cataractous images. Reformulate Eq.~\ref{eq:imaging} as
\begin{equation}
I(i,j)=\alpha\cdot L\cdot \gamma(i,j)+(1-\tau(i,j))(L-\alpha\cdot L\cdot \gamma(i,j)).
\label{eq:imaging2}
\end{equation}
Denote the ideal clear fundus image $L\cdot \gamma(i,j)$ as $s(i,j)$,
\begin{equation}
I(i,j)=\alpha\cdot s(i,j)+(1-\tau(i,j))(L-\alpha\cdot s(i,j)).
\label{eq:imaging3}
\end{equation}

As the uneven transmission function of cataract lens, a cataractous panel $J$ is employed to model $(1-\tau(i,j))$, where with the center of $(a,b)$ by $J_{ij} = \sqrt{(i-a)^{2}+(j-b)^{2}}$. Considering the disparity illumination intensity of the fundus under different spectral illuminations ~\cite{delori1977monochromatic}, decompose Eq.~\ref{eq:imaging2} into RGB channels. The cataract-like image simulation is given by:
\begin{equation}
C(s_c)=\alpha \cdot s_c\ast g_B(r_B,\sigma_B) +\beta \cdot J\ast g_L(r_L,\sigma_L)\cdot (L_c-s_c),
\label{eq:degradation}
\end{equation}
\noindent where $s_c$ is the clear image, $c\in \{r,g,b\}$ refers to the red, green or blue channel of the image, and $\ast$ denotes the convolution operator. $\alpha$ and $\beta$ denote the weight for the clear fundus image and the noise from cataracts. 
$g_B$ and $g_L$ are the Gaussian filters respectively for clear image smooth and cataractous panel, where $g(r,\sigma)$ represents a Gaussian filter with a radius of $r$ and spatial constant $\sigma$.
The highest intensity in $s_c$ is selected as $L_c$. The simulated cataract-like image is defined as $s'= [C(s_r),C(s_g),C(s_b)]$.

\begin{figure*}[htbp]
\begin{centering}
\includegraphics[width=17cm]{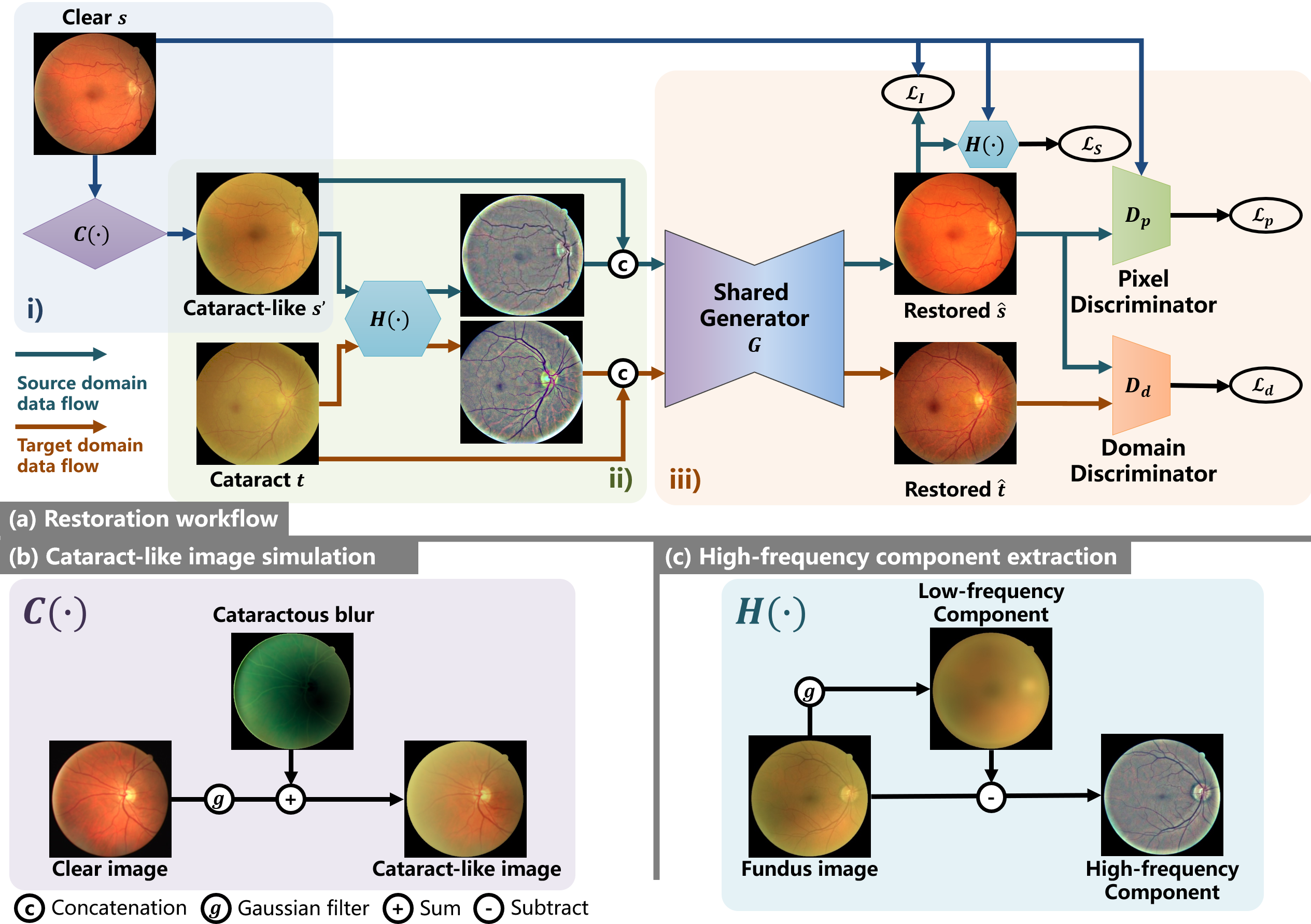}
\par\end{centering}
\caption{(a) The overall workflow of the proposed restoration algorithm. $C(\cdot )$ and $H(\cdot )$ respectively refer to the simulation of cataract-like images in (b) and the extraction of the high-frequency component in (c).
i) The cataract-like image $s'$ is simulated from clear image $s$ to synthesize the source domain of high-low quality image pairs. The target domain consists of the images to be restored. ii) Images from the source and target domain are simultaneously forwarded to the restoration network with the structure guidance of their high-frequency components. iii) $\mathcal{L}_{p}$, $\mathcal{L}_{I}$, $\mathcal{L}_{S}$ train the model to restore cataractous images, and $\mathcal{L}_{d}$ adapts the model from the source domain to the target domain.
(b) Simulating cataract-like images from clear ones following Eq.~\ref{eq:degradation}.
(c) According to Eq.~\ref{eq:lowpass}, the HFC is extracted from the fundus image as the structure guidance in the restoration. 
} \label{fig:overview}
\end{figure*}

Fig.~\ref{fig:overview} (b) demonstrates the simulation of cataracts from clear fundus images using Eq.~\ref{eq:degradation}.
The source domain of high-low quality paired data is thus formed by clear fundus images and synthesized cataract-like ones. Additionally, owing to the domain adaptation algorithm in the following part, the rigorous requirements are unnecessary for the similarity between the simulated and real cataractous ones.

\subsection{Structure guidance with HFC}
Segmentation masks have been used in previous algorithms~\cite{shen2020understanding,luo2020dehaze} to preserve retinal structures in fundus image restoration. 
A segmentation network was implemented in~\cite{shen2020understanding} to inject retinal structural information into the restoration network by feature map concatenation. 
While a structure segmentation loss was proposed in~\cite{luo2020dehaze} to preserve retinal structures, which was calculated from two pretrained U-nets.
However, this preservation requires segmentation annotation and ignores the structures beyond the segmentation masks, such as pathological characteristics, which are fundamental in the fundus assessment. To address the disadvantage of segmentation-based structure preservation, the HFC is automatically extracted from the cataractous image as the structure guidance.  

Inspired by the frequency space analysis for cross-domain tasks~\cite{yang2020fda,huang2021fsdr}, the domain-invariant features in frequency space are extracted to preserve fundus structures. Based on the retinex theory~\cite{land1977retinex}, the blur is caused by illumination bias, which is considered to be a low-frequency noise in the image. Further, as presented in Eq.~\ref{eq:imaging}, the light reflected from the cataract belongs to the low-frequency component (LFC) in the fundus image $I$, and the reflection of the retina is intact in HFC. Thus it indicates that the HFC of fundus images is more robust to cataracts and contains most domain-invariant structure features. Decompose the cataractous fundus image into two additive portions of LFC $L(I)$ and HFC $H(I)$:
\begin{equation}
I = L(I) + H(I).
\label{eq:fc}
\end{equation}

Therefore, HFC of cataractous images is employed to preserve retinal structures during the restoration. As presented in Fig.~\ref{fig:overview} (c), LFC can be extracted efficiently using a low-pass Gaussian filter, and HFC is calculated from:
\begin{equation}
H(I)= I - I\ast g_P(r_P,\sigma_P),
\label{eq:lowpass}
\end{equation}
\noindent where the low-pass Gaussian filter $g_P$ with radius $r_P$ and spatial constant $\sigma_P$ is implemented to capture LFC from image $I$. Specifically, to guarantee that entire retinal structures are contained in HFC, the kernel size of the Gaussian filter should be sufficiently large. $r_P$ was set as 26, and $\sigma_P$ was calculated as 9 according to $r_P$~\cite{getreuer2013survey}.

With the structure guidance of HFC, the segmentation is replaced and retinal structures, especially pathological characteristics are comprehensively preserved.
\subsection{Restoration with unsupervised domain adaptation}
The synthesized high-low quality paired images are leveraged to train the restoration model, and HFC preserves retinal structures while removing cataracts from fundus images. Furthermore, unsupervised domain adaptation is implemented to generalize the restoration model from the synthesized source domain to the target domain of real cataracts. 

\subsubsection{Training restoration model}
The paired image-to-image translation network with the generator of U-net is implemented to learn the restoration model from the synthesized source domain of paired images. To preserve retinal structures during the restoration, HFC is extracted from the cataract-like image $s'$. As expressed in Fig.~\ref{fig:overview}(a), $s'$ is forwarded to the generator $G$ concatenating with the HFC as the structure guidance. The restored image $\hat{s}$ is given by
\begin{equation}
\hat{s} = G(s', H(s')).
\label{eq:shat}
\end{equation}

For optimizing the generator $G$, threefold loss functions are defined to minimize. An adversarial loss $\mathcal{L}_p$ for $G$ is quantified by $D_p$, a discriminator for the pixels from the restored image $\hat{s}$ and the clear one $s$, where
\begin{equation}
\begin{aligned}
\mathcal{L}_{p}=&\mathbb{E}\left [ \mathrm{log}D_p\left ( s \right ) \right ]+\mathbb{E}\left [ \mathrm{log}\left (1-D_p  \left ( G\left ( {s}' , H({s}')\right ) \right ) \right )\right ].
\end{aligned}
\label{eq:p}
\end{equation}

L1 distance is also calculated between $\hat{s}$ and $s$, leading $\hat{s}$ to the ground truth. 
\begin{equation}
\mathcal{L}_{I}=\mathbb{E}[\left \| s-G({s}', H({s}')) \right \|_{1}].
\label{eq:lI}
\end{equation}

To guarantee that the realistic and convincing contents are properly preserved, another L1 distance is measured between the HFCs of $s$ and $\hat{s}$.
\begin{equation}
\mathcal{L}_{S}=\mathbb{E}[\left \| H(s)-H(\hat{s})) \right \|_{1}].
\label{eq:ls}
\end{equation}

Thus the restoration model is learned on the synthesized image pairs from the source domain by minimizing the loss function $\mathcal{L}_{G}$, where
\begin{equation}
\mathcal{L}_{G}=\mathcal{L}_{p}+\lambda_1 \mathcal{L}_I+\lambda_2 \mathcal{L}_S,
\label{eq:G}
\end{equation}

\subsubsection{Unsupervised domain adaptation}
The effectiveness of the restoration model strongly relies on an assumption, that the training and test data are independent and identically distributed. However, since the images for training are synthesized and the images to be restored are collected from clinics, the domain shift between the training and test data is inevitable in this study.
Therefore, to effectively deploy the restoration model, unsupervised domain adaptation is introduced to transfer the restoration model from the synthesized to real images.

The unsupervised domain adaptation is implemented based on the outputs of the restoration model in the absence of annotation. 
As the aim of this study is to restore clear fundus images, our intuition is that properly restored images should belong to an identical domain no matter the cataract images are from the source or target domain. 
Consequently, as demonstrated in Fig.~\ref{fig:overview} (a) , another discriminator for domains, $D_d$, is adopted in the output space to conduct an unsupervised domain adaptation.

Specifically, the real cataractous images are employed as the target domain. The source domain and the target domain are simultaneously forwarded to the restoration model. The two domains share the generator $G$, and a discriminator $D_d$ is used to import adversarial learning for domain adaptation. The loss of domain adaptation is subsequently calculated from the restoration result $\hat{s}$ of the source domain and $\hat{t}$ of the target one. 
The loss function is formulated as:
\begin{equation}
\begin{aligned}
\mathcal{L}_{d}=\mathbb{E}\left [ \mathrm{log}D_d\left (\hat{s} \right ) \right ]+\mathbb{E}\left [ \mathrm{log}\left (1-D_d  \left (\hat{t}\right ) \right ) \right ],
\end{aligned}
\label{eq:dgan}
\end{equation}
\noindent where $\hat{t}=G\left ( t,H(t) \right )$, and $t$ represents real cataract fundus images. 
Append $\mathcal{L}_{d}$ into $\mathcal{L}_{G}$. During the training stage, to fool discriminator $D_d$, and the adversarial loss $\mathcal{L}_{d}$ leads $G$ to generate $\hat{s}$ and $\hat{t}$, which shall share an identical distribution. Therefore, the restoration model adapts to the real cataractous images without supervision.

Finally, based on the described modules, the annotation-free restoration network for cataractous images is trained without any annotations using the loss function 
\begin{equation}
\mathcal{L}_{total}=\mathcal{L}_{p}+\lambda_1 \mathcal{L}_{I}+\lambda_2 \mathcal{L}_S+\lambda_3 \mathcal{L}_d,
\label{eq:loss}
\end{equation}
\noindent where $\lambda_1$, $\lambda_2$ and $\lambda_3$ are the weights used to balance the losses and were respectively set to 100, 50 and 1 in this paper.

\subsection{Network implementation}
ArcNet adopts an Unet-like network as a generator and two patch-based fully convolutional networks as discriminators. The generator consists of 8 down-sampling layers and 8 up-sampling layers with skip connections between the corresponding layers. Each down-sampling layer contains a down convolution layer, a leaky ReLU layer, and a batch normalization layer. While each up-sampling layer contains a transposed convolution layer, a ReLU layer, and a batch normalization layer, and the output layer contains a ReLU layer, a convolution layer, and a Tanh function. The pixel discriminator and domain discriminator are with the same network architecture which adopts 5 convolutional layers with kernel size 4 $\times$ 4 and stride of $\{$2, 2, 2, 1, 1$\}$ for the channel number of $\{$64, 128, 256, 512, 1$\}$.

In the training phase, the input image size was set to 256 $\times$ 256 and the batch size was set to 8. 
To reduce the gap between the source domain and the target domain, the model loaded images randomly in a scale, including 286, 306, 326, and 346. After which it randomly cropped the images to the size of 256 $\times$ 256.
All the networks were trained for 100 epochs using Adam optimizer with a learning rate equal to $2\times 10^{-4}$ for the 80 epochs and $5\times 10^{-5}$ for the next 20 epochs. During inference, the input image size was set to 256 × 256, and the batch size was set to 1.

The network was implemented in PyTorch, and was trained on an NVIDIA TITAN RTX with 24GB of memory. The number of model parameters is about 60 million.

\begin{figure*}[htbp]
\begin{centering}
\includegraphics[width=16cm]{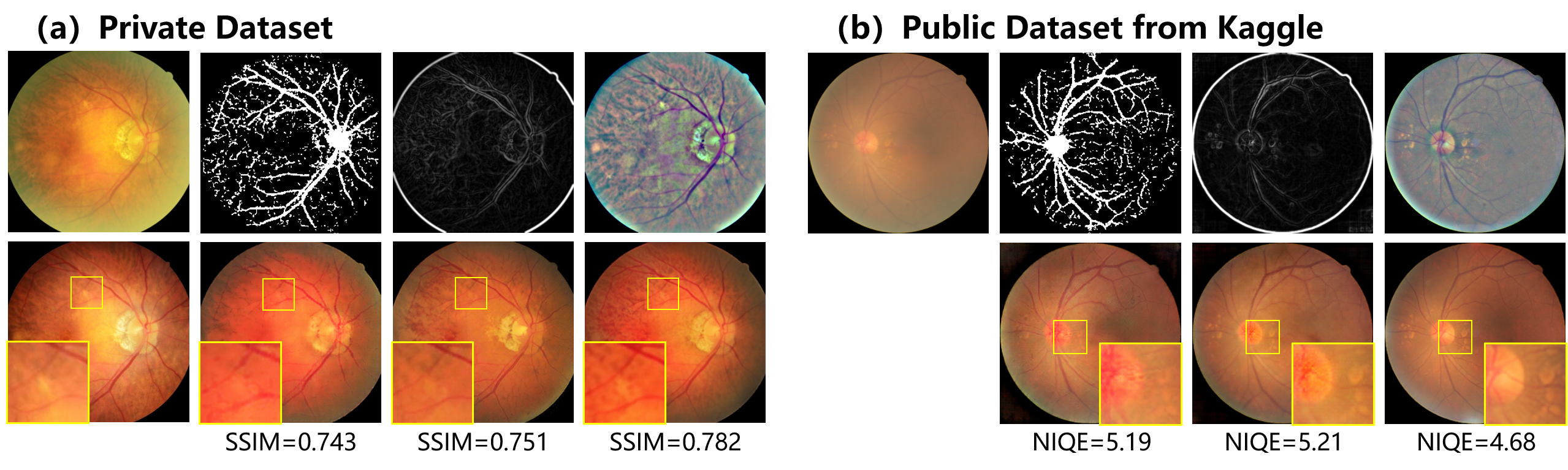}
\par\end{centering}
\caption{Restoration guided by segmentation, Sobel operator, and HFC. (a) Private dataset. (b) Public cataract subset of Kaggle. The images in the first row are successively the cataractous image, the structure guidance from segmentation, Sobel operator, and HFC. The second row exhibits the ground truth of clear fundus image (unavailable for Kaggle), the results of the restoration with the guidance of segmentation, Sobel operator, and HFC. 
The yellow boxes highlight pathological characteristics of drusen.} \label{fig:guidance}
\end{figure*}

\section{Experiments}
Five experiments were implemented to interpret the performance of ArcNet. The effectiveness of HFC and the proposed modules were first demonstrated. Then comparisons with state-of-the-art algorithms were implemented to verify ArcNet on the restoration of cataractous images. Subsequently, an investigation was provided to understand the impact on restoration from selecting source domains. An application of ArcNet was then presented to promote the diagnosis of ocular fundus diseases. 
Finally, the model setting and computational complexity of ArcNet were analyzed.

\subsection{Experiment Setting}
\subsubsection{Datasets}
Four fundus image datasets for two tasks were employed in the experiments to comprehensively verify the proposed ArcNet. The experimental protocol is in accordance with the Declaration of Helsinki and was approved by the local Ethics Committee.

As provided in Table~\ref{tab:dataset}, the evaluation of restoration performances were conducted using three datasets: 1) a private dataset, RCF, consisting of 52 paired fundus images collected before and after cataract surgery from Peking University Third Hospital; two public fundus datasets were collected from 2) DRIVE\footnote[1]{http://www.isi.uu.nl/Research/Databases/DRIVE/} and 3) Kaggle\footnote[2]{https://www.kaggle.com/jr2ngb/cataractdataset}.
To exhibit the effectiveness of ArcNet in diagnosis tasks, 4) the private ophthalmic dataset, Fundus-iSee, is used to diagnose ocular comorbidities in cataract patients.

\begin{table}[htb]
\footnotesize
\centering {\caption{Datasets used in the experiments}
\label{tab:dataset} }
\renewcommand{\arraystretch}{1.2}
\begin{tabular}{|c|p{3cm}| p{3cm} |}
\hline
Tasks & Target domain datasets& Source domain datasets \\
\hline
\multirow{2}{*}{Restoration} & 1) RCF: 26  images before surgery paired with 26 after surgery & 2) DRIVE: 40 clear images with segmentation annotations \\
& 3-1) Cataractous subset of Kaggle: 100 cataractous images & 3-2) Normal subset of Kaggle: 300 clear images \\
\hline
Diagnosis & \multicolumn{2}{p{6cm}|}{4) Fundus-iSee: 10000 fundus images from patients with (2669) and without  (7331) cataracts are sorted into five categories according to fundus diseases} \\
\hline
\end{tabular}
\end{table}

\subsubsection{Baseline}
\label{baseline}
Seven state-of-the-art restoration algorithms for fundus images were conducted as the baseline. Three hand-crafted methods based on CLAHE~\cite{mitra2018enhancement}, SGRIF~\cite{cheng2018structure}, and image filtering~\cite{cao2020retinal}, were leveraged to enhance the fundus images without training. In contrast, the image-to-image translation networks of pix2pix \cite{isola2017image} and CycleGAN \cite{zhu2017unpaired} were trained on the source domain and tested on the target domain to restore cataract images.  Moreover, the recent restoration approaches~\cite{shen2020understanding,luo2020dehaze} using segmentation annotations were also performed. Additionally, the source and target domain without annotations were forwarded to ArcNet during the training stage. Then the learned model was applied to restore the cataract images of the target domain.

\subsubsection{Evaluation metrics}
The annotation is only available in the private dataset because of the difficulties in collecting high-low quality paired data. Thus different evaluation metrics are selected to quantify the restoration performance according to the attendance and absence of the ground truth.
For the annotated private dataset, the evaluation metrics of structural similarity (SSIM) and peak signal to noise ratio (PSNR) were used to quantify the restored image quality. 
While for the unannotated public dataset, the evaluation metrics of image quality, i.e., inception scores (IS) and natural image quality evaluator (NIQE)~\cite{guo2020study, zhao2019data}, were adopted to assess the restoration performance.

Particularly, registration was applied between the paired images in the private dataset, and SSIM and PSNR were calculated from the overlap area of the paired images. The PSNR and SSIM values grow with an increase in the similarity between the restoration result and the ground truth. IS and NIQE quantify the image quality by evaluating the distortion, and the satisfying image quality is indicated by the high IS score and the low NIQE score.

\subsection{Structure guidance for ArcNet}
In previous restoration algorithms, structure segmentation and edge detection were imported to preserve retinal structures.
In~\cite{shen2020understanding} and \cite{luo2020dehaze}, segmentation networks were leveraged to provide retinal structural information and loss for restoration.
Alternative to segmentation networks, the edge detected by the Sobel operator was utilized in our previous study~\cite{li2021Restoration} with an architecture same to ArcNet to preserve retinal structures during the restoration.

To demonstrate the advantage of the HFC and compare the proposed algorithm with the previous one~\cite{li2021Restoration}, in this section, ArcNet is respectively implemented with structure guidances of segmentation, Sobel operator, and HFC.
Specifically, a U-net was trained on DRIVE with the segmentation annotations of vessels and optic disks. Thus the HFC is replaced by segmentation masks captured by the U-net as the structure guidance in ArcNet. And as our previous algorithm~\cite{li2021Restoration} shares an identical architecture with ArcNet, it is also presented and compared in the restoration with the Sobel operator.

The restoration was respectively conducted with the source-target domain groups of DRIVE and RCF, together with the normal and cataract subset of Kaggle. 
Fig.~\ref{fig:guidance} exhibits the restoration results as well as the guidance of segmentation, Sobel operator, and HFC.
Pathological characteristics of drusen are highlighted in the yellow boxes of the second row.
As observed from the first row, the segmentation guidance only attends to the annotated structures, such that the unannotated pathological characteristics were lost in the second row. Moreover, due to the quality degradation caused by cataracts, the guidance suffers from segmentation errors.
Though the Sobel operator is sensitive to the edge and forwards more pathological characteristics to the network, it leads our previous algorithm~\cite{li2021Restoration} to ignore region information. Thus artifacts appeared in the restored optic disk of Fig.~\ref{fig:guidance} (b).
HFC preserves most details of the retinal structures, such as the tiny particulars of the optic disc and retina. Consequently, the structures distorted in the restoration guided by segmentation and Sobel operator, were desirably preserved in the image restored with HFC.

\begin{table}[htb]
\footnotesize
\centering {\caption{Comparison on restoration results with DRIVE and Kaggle.}
\label{tab:guidance} }
\renewcommand{\arraystretch}{1.2}
\begin{tabular}{c p{0.9cm} p{0.93cm} c p{0.9cm} p{0.9cm}}
\hline
\multirow{2}{*}{Guidance}& \multicolumn{2}{c}{DRIVE} &&\multicolumn{2}{c}{Kaggle}\\
\cline{2-3}\cline{5-6}
 &SSIM $\uparrow$ & PSNR $\uparrow$  &&IS $\uparrow$ & NIQE $\downarrow$ \\
\hline
Segmentation & 0.738 & 18.05 && 1.46 & 5.12\\ 
Sobel~\cite{li2021Restoration}& 0.755 & 18.07 && 1.50 & 5.32\\
HFC (ours)& \textbf{0.767} & \textbf{18.29} && \textbf{1.52} & \textbf{4.52}\\
\hline
\end{tabular}
\end{table}

The quantitative comparison is summarized in Table~\ref{tab:guidance}. The evaluation metrics of SSIM and PSNR were calculated for the private dataset, and the metrics of IS and NIQE assessed the restoration of Kaggle.
Due to the capacity of ArcNet, all structure guidances achieve decent scores in the evaluation metrics. Furthermore, superior performance was displayed using the proposed guidance of HFC. 
More meaningful details are forwarded to the network by HFC without annotation.
Therefore, it indicates that HFC is more efficient in preserving the retinal structures during the restoration. 
\subsection{Ablation study of ArcNet}

For illuminating the effectiveness of the modules in ArcNet, a comparison against the ablation of modules was conducted.DRIVE and RCF were still used as the source and the target domain here. 
The fundamental framework of ArcNet follows pix2pix~\cite{isola2017image}, which in Fig.~\ref{fig:overview} includes the generator $G$, the adversarial loss $\mathcal{L}_p$ of the discriminator $D_p$, as well as the L1 distance $\mathcal{L}_I$. 
Subsequently, the proposed modules refer to: a) $H(\cdot)$ introduces the structure guidance of HFC to the restoration network; b) the structure loss $\mathcal{L}_S$ emphasizes structure preservation in the restoration; c) the domain loss $\mathcal{L}_d$ of the discriminator $D_d$ achieves unsupervised domain adaptation between the source and target domain in ArcNet.

\begin{figure}[htbp]
\begin{centering}
\includegraphics[width=7cm]{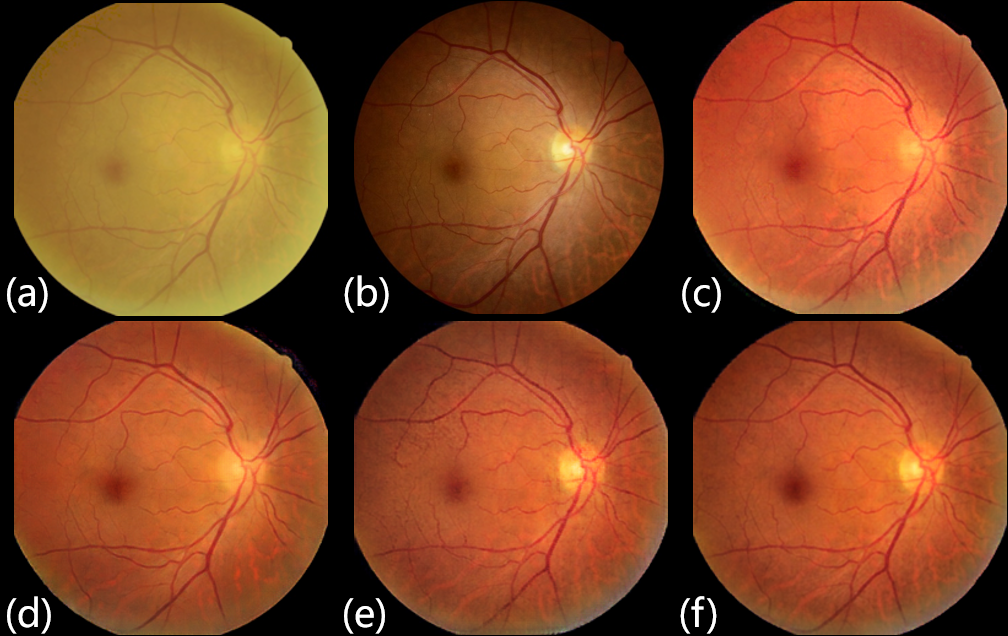}
\par\end{centering}
\caption{The effect of the proposed modules. (a) cataract image. (b) the corresponding clear image (c) The fundamental network. (d) w/ $H(\cdot)$, HFC is imported to the restoration. (e) w/ $H(\cdot)$ and $\mathcal{L}_S$, structure preservation is emphasized. (f) w/ $H(\cdot)$, $\mathcal{L}_S$, and $\mathcal{L}_d$, unsupervised domain adaptation is further achieved in ArcNet. Zoom in for details.} \label{fig:ablation}
\end{figure}

\begin{figure*}[htbp]
\begin{centering}
\includegraphics[width=15cm]{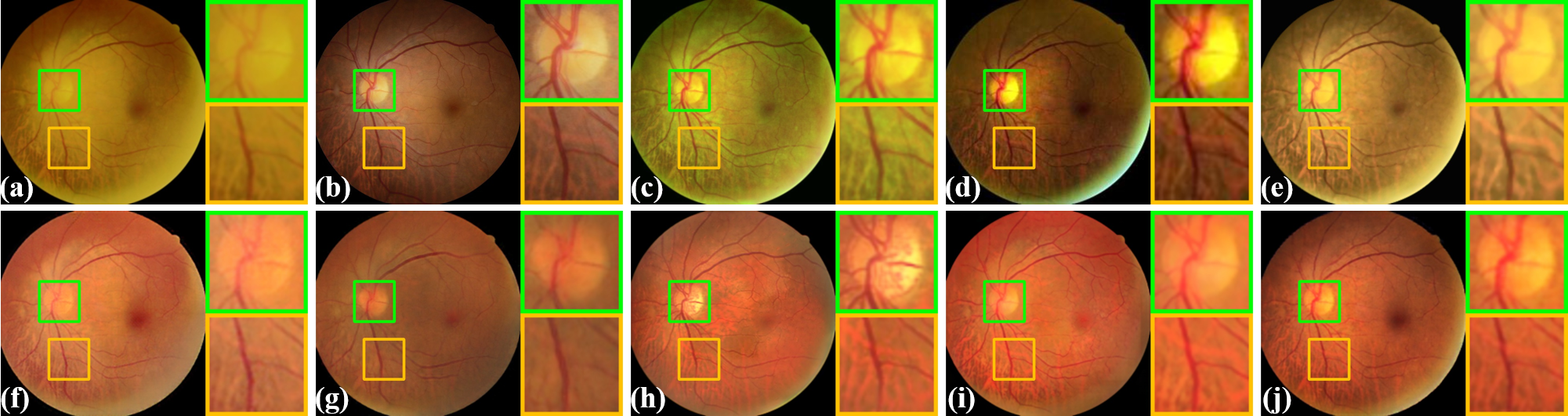}
\par\end{centering}
\caption{Visual comparison of images restored from cataract ones. (a) cataract image. (b) clear fundus image after surgery. (c) Mitra et al.~\cite{mitra2018enhancement}. (d) SGRIF~\cite{cheng2018structure}. (e) Cao et al.~\cite{cao2020retinal}. (f) pix2pix \cite{isola2017image}. (g) CycleGAN \cite{zhu2017unpaired}. (h) Luo et al. \cite{luo2020dehaze}. (i) CofeNet \cite{shen2020understanding}. (j) ArcNet (ours).} \label{fig:private}
\end{figure*}

\begin{figure*}[htbp]
\begin{centering}
\includegraphics[width=15cm]{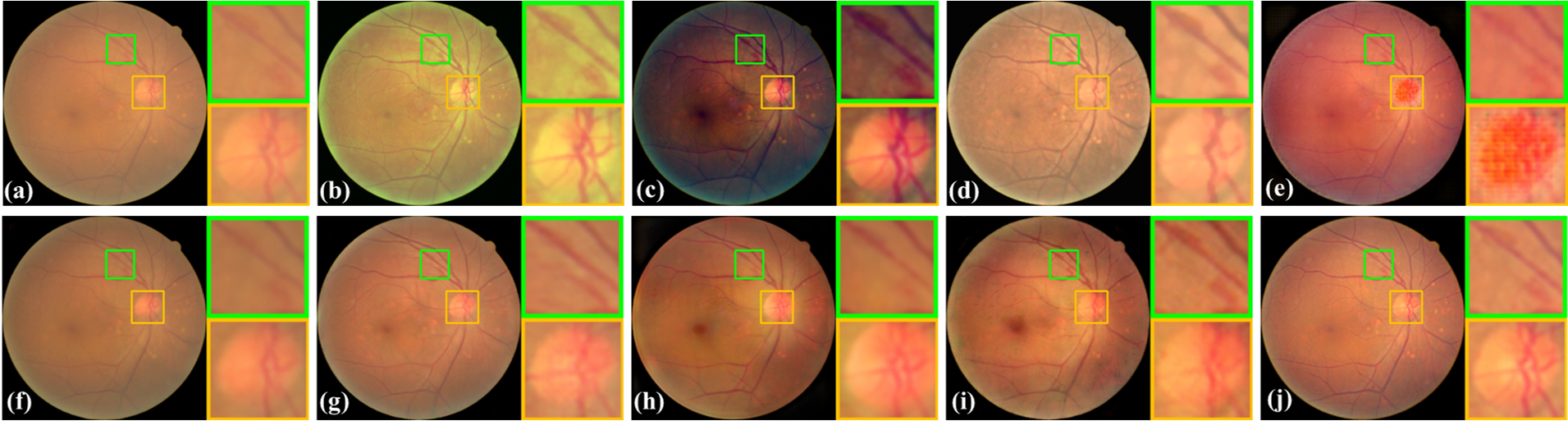}
\par\end{centering}
\caption{Visual comparison of results for the public Kaggle dataset. (a) cataract image. (b) Mitra et al.~\cite{mitra2018enhancement}. (c) SGRIF~\cite{cheng2018structure}. (d) Cao et al.~\cite{cao2020retinal}. (e) pix2pix \cite{isola2017image}. (f) CycleGAN \cite{zhu2017unpaired}. (g) CycleGAN-real. (h) Luo et al. \cite{luo2020dehaze}. (i) CofeNet \cite{shen2020understanding}. (j) ArcNet (ours).} \label{fig:public}
\end{figure*}

The ablation study was conducted by individually deploying the modules to the fundamental network. $H(\cdot)$ was deployed first to import HFC, and then $\mathcal{L}_S$ promoted structure preservation, finally domain adaptation was presented by $\mathcal{L}_d$.
Fig.~\ref{fig:ablation} visualizes the results of the restoration, and quantitative evaluation is summarized in Table~\ref{tab:Ablation}.

Due to unsupervised restoration for the target domain,  the gap between the intensity distributions of the ground truth (Fig.~\ref{fig:ablation} (b)) and the restored image (Fig.~\ref{fig:ablation} (c-f)) was inevitable. Therefore, our observation and analysis concentrate on content consistency.
Compared with the result in Fig.~\ref{fig:ablation} (c) restored by the fundamental network, HFC improves the preservation of retinal structures in Fig.~\ref{fig:ablation} (d). In Fig.~\ref{fig:ablation} (e), $\mathcal{L}_S$ further enhances the structures, especially in the optic disk area. 
Finally, in the comparison between Fig.~\ref{fig:ablation} (e) and (f), domain adaptation by $\mathcal{L}_d$ promises ArcNet a restoration result with realistic textures and appearance. 
Moreover, the quality of the restored images gradually increased in Table~\ref{tab:Ablation} further demonstrating the proposed modules' effectiveness.

\begin{table}[ht]
\footnotesize
\centering {\caption{\deleted{Ablation study of the proposed network.}\added{Ablation study. Our full model achieves reasonable results in both metrics.}}
\label{tab:Ablation} }
\renewcommand{\arraystretch}{1.2}
\begin{tabular}{p{0.7cm} p{0.7cm} p{0.7cm} ||p{0.1cm} p{1.3cm} p{1.3cm}}
\hline
$H(\cdot)$ & $\mathcal{L}_S$ & $\mathcal{L}_d$ && SSIM $\uparrow$ & PSNR $\uparrow$ \\
\hline
& & && 0.733 & 17.69 \\
$\surd$ & &  &&  0.755 & 17.83 \\
$\surd$ & $\surd$ & &&  0.763 & 17.96 \\
$\surd$ & $\surd$ & $\surd$ && \textbf{0.767} & \textbf{18.29}  \\
\hline
\end{tabular}
\end{table}

\subsection{Comparison with existing methods}
To verify the advantages of ArcNet, restoration comparisons on cataractous images were conducted in this section. The comparisons were implemented on the private dataset with ground truth, as well as the unannotated public datasets. The state-of-the-art methods described in Section~\ref{baseline} were introduced as the baselines.

\subsubsection{Restoration of private dataset}
The source domain of DRIVE and the target domain of RCF were employed to conduct this experiment.
The visualized and quantitative comparisons against existing methods were summarized in Fig. \ref{fig:private} and Table \ref{tab:private}. The proposed ArcNet achieves outstanding performance, and as shown in Fig. \ref{fig:private} the fundus images restored by ArcNet appear relatively realistic and clean details. Additionally, consistent with Table \ref{tab:Ablation}, in Table \ref{tab:private} the guidance of HFC and domain adaptation endows ArcNet with particularly favorable performance against the existing methods.

\begin{table}[htb]
\footnotesize
\centering {\caption{\deleted{Comparison with existing methods on private data.}\added{Quantitative comparison with state-of-the-art methods on the private dataset.  Our method produces superior results in the metrics with reference.}}
\label{tab:private} }
\renewcommand{\arraystretch}{1.2}
\begin{tabular}{c p{0.4cm} p{1.4cm} p{1.4cm}}
\hline
Algorithms &&SSIM $\uparrow$ & PSNR $\uparrow$ \\ 
\hline
Mitra et al.~\cite{mitra2018enhancement}&& 0.698 & 15.82 \\
SGRIF~\cite{cheng2018structure}&& 0.609  & 15.07 \\
Cao et al.~\cite{cao2020retinal}&& 0.709  & 16.40 \\
pix2pix \cite{isola2017image}&& 0.731 & 17.68 \\
CycleGAN \cite{zhu2017unpaired}&& 0.732 & 17.25 \\
Luo et al. \cite{luo2020dehaze}  && 0.704 & 17.29 \\
CofeNet \cite{shen2020understanding} && 0.754 & 18.03 \\
ArcNet (ours)&& \textbf{0.767} & \textbf{18.29} \\
\hline
\end{tabular}
\end{table}

It can be observed from the results that the hand-crafted methods~\cite{mitra2018enhancement,cheng2018structure,cao2020retinal} fail to restore the fundus images in the appearance of ground truth. These methods are manually designed according to publicly distributed attributes of hazed fundus images, which often results in limited versatility to noise distributions. 
In contrast, the images restored using the methods of image-to-image translation \cite{isola2017image,zhu2017unpaired}, appear similar to the clear images. The paired training data allow image-to-image translation models to explore a particular mapping between the cataract-like and clear images. 
Segmentation masks were introduced to preserve retinal structures in~\cite{shen2020understanding,luo2020dehaze}, and thus they are more sensitive to the vessels and optic disks.

However, the inherent difference between the cataract-like images from the source domain and real cataract images from the target domain leads the methods to alias the retinal structures, such that the details became blurry and of low contrast. 
ArcNet is capable of mapping the appearance of clear fundus images and enhancing the fine retinal structures.

\subsubsection{Restoration of public data}
The comparison is also performed on a dataset of a large volume to examine the proposed method. The dataset from Kaggle was employed to conduct this experiment, where the source domain was synthesized from the normal subset, and the cataractous subset was used as the target domain. Due to the absence of clear images paired to the cataractous ones, the results were assessed using the evaluation metrics of IS and NIQE without the reference of ground truth.

\begin{table}[htbp]
\footnotesize
\centering {\caption{\deleted{Comparison with existing methods on public data.}\added{Quantitative comparison with state-of-the-art methods on the public dataset. Superior results are presented by our method in the metrics without reference.}}
\label{tab:public} }
\renewcommand{\arraystretch}{1.2}
\begin{tabular}{c p{0.4cm} p{1.4cm} p{1.4cm}}
\hline
Algorithms && IS $\uparrow$ & NIQE $\downarrow$\\
\hline
Mitra et al.~\cite{mitra2018enhancement}&& 1.31 & 6.44 \\
SGRIF \cite{cheng2018structure}&& 1.44 & 5.98 \\
Cao et al.~\cite{cao2020retinal}&& 1.45  & 5.98 \\
pix2pix \cite{isola2017image}&& 1.49 & 6.30 \\
CycleGAN \cite{zhu2017unpaired}&& 1.40 & 5.25 \\
CycleGAN-real \cite{zhu2017unpaired}&& 1.47 & 5.30 \\
Luo et al. \cite{luo2020dehaze} && 1.45 & 5.56 \\
CofeNet \cite{shen2020understanding} && 1.51 & 5.49 \\
ArcNet (ours) && \textbf{1.52} & \textbf{4.52} \\
\hline
\end{tabular}
\end{table} 
Fig.~\ref{fig:public} exhibits the results of the restoration. And as plenty of unpaired normal and cataractous images were contained in the Kaggle dataset, CycleGAN was not only trained on the synthesized data, but also on the real unpaired data (denoted as CyclaGAN-real). 
The optic disk and a retinal hemorrhage are highlighted to demonstrate the capacity of retinal structure preservation. 
The structures are properly preserved by ~\cite{mitra2018enhancement,cheng2018structure}, but the image is mapped to an inferior color and illumination space. In the results of \cite{shen2020understanding,cao2020retinal}, the retinal hemorrhage is preserved, while the vessels in the optic disk are fuzzy. Compared with the existing algorithms, ArcNet satisfactorily preserved the optic disk and retinal hemorrhage, as well as mapped the image to proper image space. 

The metrics of IS and NIQE provides quantitative evaluations. The quality of restoration results is summarized in Table~\ref{tab:public}. The quantitative evaluations in Table~\ref{tab:public} further demonstrate that ArcNet achieves reasonable and obvious improvements of the restoration.

\subsection{Selection of source domain}
Since the restoration model was learned from the source domain, it can infer that the selection of the source domains impacts the restoration performance. To understand the impact of the selection of source domains, an investigation was conducted by comparing the images restored using different source domains. The dataset of DRIVE, and the normal subset from Kaggle were respectively employed to simulate cataract-like images and synthesize the source domain. And RCF was used as the target domain.

\begin{figure}[hbt]
\begin{centering}
\includegraphics[width=7cm]{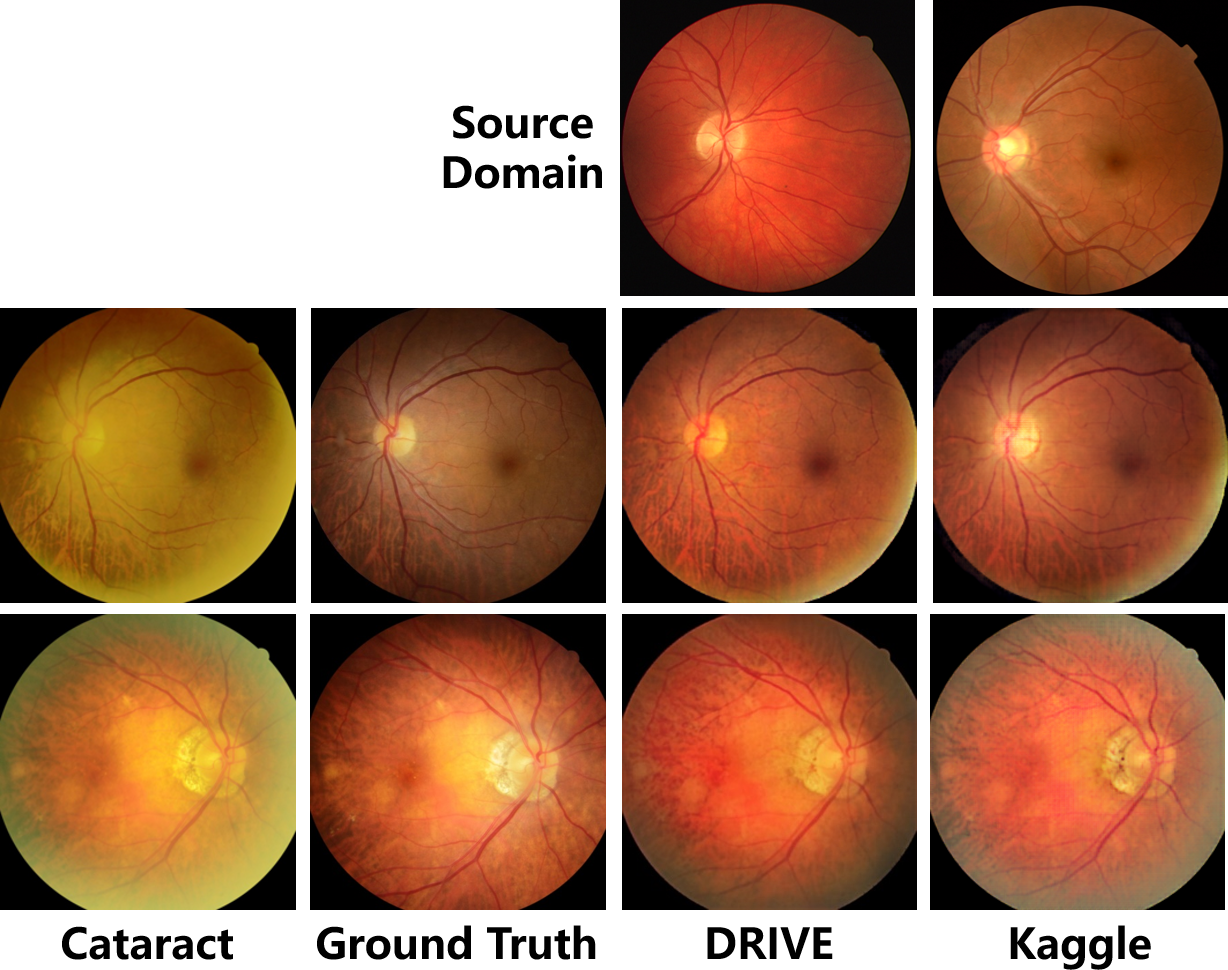}
\par\end{centering}
\caption{Images were restored using different source domains. The first row exhibits the examples from DRIVE and Kaggle, which are used as the source domain to restore cataractous images. The following rows provide the restoration results using the corresponding source domain.} \label{fig:source}
\end{figure}

The images restored using different source domains are visualized in Fig. \ref{fig:source}. The impact of the source domain was observed from the restored images. Because the images in DRIVE have a higher brightness in red than those in Kaggle, the higher brightness in red was observed from the images restored using DRIVE compared to Kaggle.
Due to the training on the source domain and the alignment between the source domain's clear images and the target domain's outputs, it is no surprise that the learned model tries to improve the quality of cataractous images with the appearances of the source domain.
Alternatively, due to the high image quality of DRIVE and Kaggle, the images restored using both source domains by the proposed ArcNet always achieve a remarkable quality. 

The quantitative comparison of the restoration using \added{the target domain of RCF with} the source domain of DRIVE and Kaggle is presented in Table~\ref{tab:source}, and the performances of the existing CNN-based algorithms were summarized as well.
The hand-crafted methods were omitted in this experiment, since their performances are independent of the source domain. 
Compared to the existing algorithms, the proposed method stably achieves the best image quality with both source domains.

\begin{table}[htbp]
\footnotesize
\centering {\caption{Comparison on restoration with the source domain of DRIVE and Kaggle.}
\label{tab:source} }
\renewcommand{\arraystretch}{1.2}
\begin{tabular}{c p{0.8cm} p{0.8cm} c p{0.8cm} p{0.8cm}}
\hline
\multirow{2}{*}{Algorithms}& \multicolumn{2}{c}{DRIVE} &&\multicolumn{2}{c}{Kaggle}\\
\cline{2-3}\cline{5-6}
 &SSIM & PSNR  &&SSIM & PSNR \\
\hline
pix2pix \cite{isola2017image}& 0.731 & 17.68 && 0.739 & 17.71\\
CycleGAN \cite{zhu2017unpaired}& 0.732 & 17.25 && 0.727 & 17.40\\ 
Luo et al. \cite{luo2020dehaze} & 0.704 & 17.29 && 0.735 & 17.74\\
CofeNet \cite{shen2020understanding} & 0.754 & 18.03 && 0.759 & 18.54\\
\added{Previous work~\cite{li2021Restoration}} & 0.755 & 18.07 && 0.748 & 18.24\\
ArcNet (ours)& \textbf{0.767} & \textbf{18.29} && \textbf{0.760} & \textbf{18.59}\\ 
\hline
\end{tabular}
\end{table}
This experiment confirms the impact on restoration from the selection of the source domain. The intensity distribution of restored images follows the source domain. Fortunately, despite the discrepancy in color, the readability and quality of cataractous images could be improved once the source domain of high-quality is employed. And the outstanding performance of ArcNet is validated using different source domains. 

\subsection{Clinical Image Restoration and Diagnosis}
Considering that the restoration of cataractous fundus images was conducted to promote the clinical observation and diagnosis, ArcNet was imported to a diagnosis task of fundus ocular diseases and compared with the state-of-the-art methods. The private ophthalmic dataset Fundus-iSee was used to conduct automatic diagnosis.
Fundus-iSee consists of 720 (492 clear VS 228 cataractous) images of age-related macular degeneration (AMD), 270 (181 VS 89) of diabetic retinopathy (DR), 450 (312 VS 138) of glaucoma, 790 (478 VS 312) of high myopia, and 7770 (5868 VS 1902) normal fundus. 
Because cataracts distort the observation of the retinal fundus and increase the risk of misdiagnosis, restoration algorithms were applied to boost the diagnosis. 

\begin{figure*}[htbp]
\begin{centering}
\includegraphics[width=15cm]{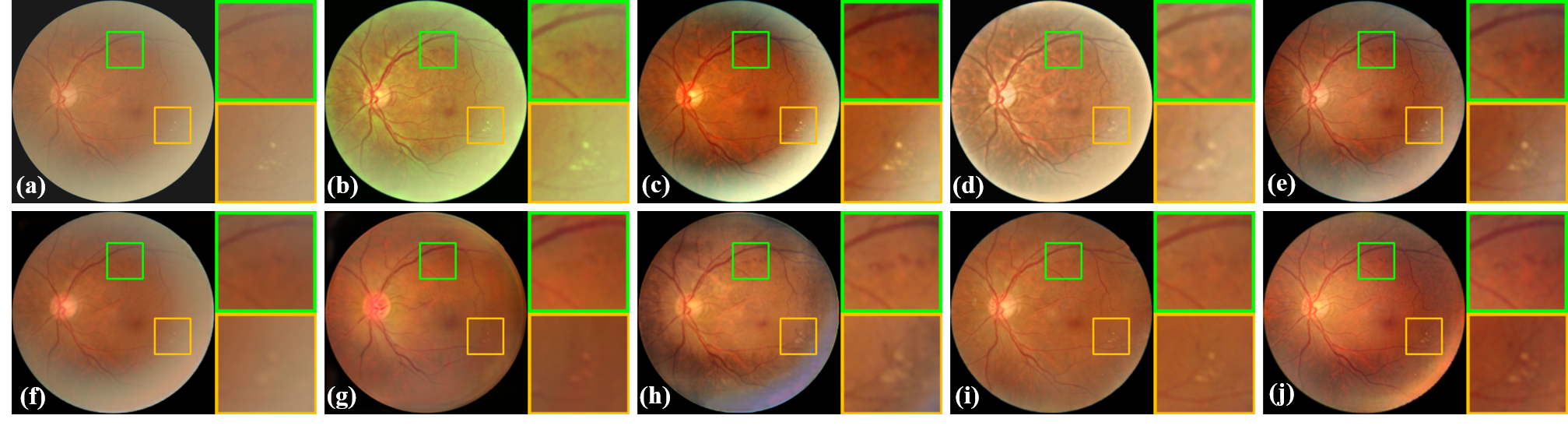}
\par\end{centering}
\caption{\added{Restoration example of the lesions in the fundus image with DR. (a) cataract image. (b) Mitra et al.~\cite{mitra2018enhancement}. (c) SGRIF~\cite{cheng2018structure}. (d) Cao et al.~\cite{cao2020retinal}. (e) pix2pix \cite{isola2017image}. (f) CycleGAN \cite{zhu2017unpaired}. (g) Luo et al. \cite{luo2020dehaze}. (h) CofeNet \cite{shen2020understanding}. (i) Previous work~\cite{li2021Restoration}. (j) ArcNet (ours).}}
\label{fig:lesion}
\end{figure*}

\begin{figure}[hbt]
\begin{centering}
\includegraphics[width=8.6cm]{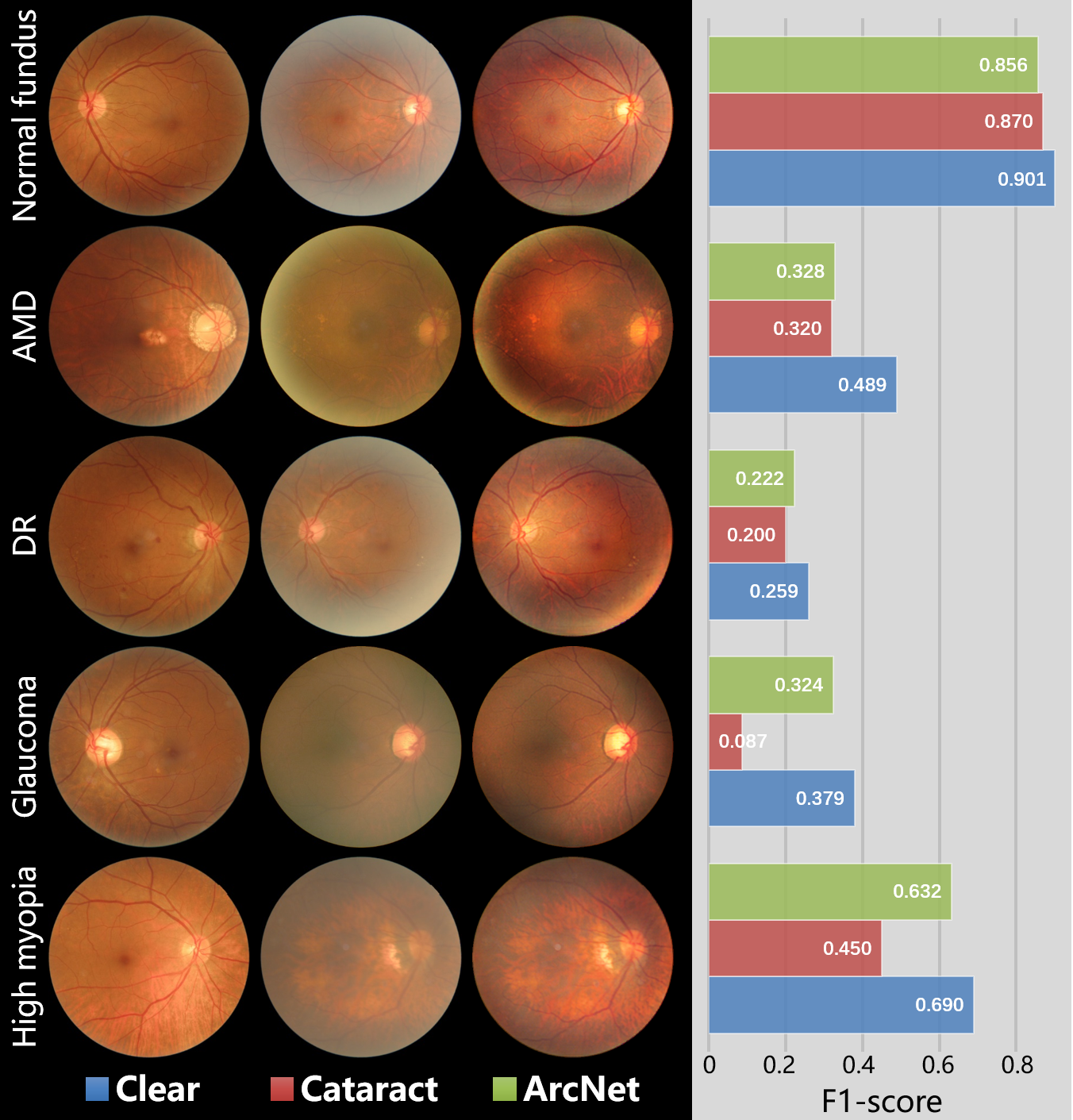}
\par\end{centering}
\caption{Samples of Fundus-iSee and the diagnosis performance with ArcNet. 
The diagnosis F1-scores of each category are presented.} \label{fig:isee}
\end{figure}

The diagnosis model was learned by ResNet-50 from 5000 randomly sampled training images. The rest 5000 images in Fundus-iSee were sorted into a clear and a cataractous subset for testing. Additionally, restoration was conducted on the cataractous subset by the state-of-the-art methods, where the restoration models were trained with the source and target domain composed of 2000 clear and 700 cataractous images from Fundus-iSee. 
\added{Fig.~\ref{fig:lesion} shows an example of the restored images with lesions of DR.}
The diagnosis results on the clear, cataractous, and restored subset are exhibited in Fig.~\ref{fig:isee}, while a comparison with the state-of-the-art methods is summarized in Table~\ref{tab:isee}. 

\added{The restoration comparison for the lesions of DR is exhibited in Fig.~\ref{fig:lesion}. As a result of cataracts, the retinal hemorrhages (green box) and hard exudates (yellow box) are obscure in the original image. 
For the state-of-the-art methods,  the lesions are diminished in the images restored by Cao et al.~\cite{cao2020retinal}, CycleGAN \cite{zhu2017unpaired}, Luo et al. \cite{luo2020dehaze},  and our previous work~\cite{li2021Restoration}, 
while Mitra et al.~\cite{mitra2018enhancement}, SGRIF~\cite{cheng2018structure}, pix2pix \cite{isola2017image}, and CofeNet \cite{shen2020understanding} preserve the lesions but result in limited image quality.
ArcNet properly restores the image, and desirably enhances the lesions of retinal hemorrhages and hard exudates.}

The image samples and F1-scores of each category are respectively presented in Fig.~\ref{fig:isee}. 
Compared to the clear images, cataracts obstruct the observation of the retinal fundus. ArcNet improves the readability of the cataractous fundus images such that fundus abnormalities are more clearly noticed in restored ones. Therefore, the diagnosis performance of ocular diseases has been remarkably boosted by ArcNet.
Moreover, due to data imbalance, the diagnosis model tends to sort any uncertain images into the category of the normal fundus. And thus, diagnosis improvement was not observed in the F1-score of the restored normal fundus images, while the advance in Ckappa verified the improvement by the restoration.

\begin{table}[htbp]
\footnotesize
\centering {\caption{Application of Restoration in Diagnosis.}
\label{tab:isee} }
\renewcommand{\arraystretch}{1.2}
\begin{tabular}{c p{0.4cm} p{1.4cm} p{1.4cm}}
\hline
Fundus Image && F1-score  & Ckappa \\
\hline
Clear && 0.838 & 0.448 \\
Cataract  && 0.730 & 0.310 \\
\hline
\hline
Mitra et al.~\cite{mitra2018enhancement}&& 0.722 & 0.364 \\
SGRIF \cite{cheng2018structure}&& 0.760 & 0.420 \\
Cao et al.~\cite{cao2020retinal}&& 0.667  & 0.326 \\
pix2pix \cite{isola2017image}&& 0.713 & 0.381 \\
CycleGAN \cite{zhu2017unpaired}&& 0.724 & 0.286 \\
Luo et al. \cite{luo2020dehaze} && 0.712 & 0.370 \\
CofeNet \cite{shen2020understanding} && 0.754 & 0.416 \\
\added{Previous work~\cite{li2021Restoration}} && 0.747 & 0.405 \\
ArcNet (ours) && \textbf{0.761} & \textbf{0.428} \\
\hline
\end{tabular}
\end{table}

A diagnosis comparison with images restored by state-of-the-art methods is summarized in Table~\ref{tab:isee}, which is evaluated by the metrics of F1-score and Cohen's kappa (Ckappa).
It should be noted that Ckappa is more sensitive than F1-score in multi-categories classification, as it has a more realistic view of the model’s performance when using imbalanced data~\cite{jeni2013facing}.
According to the diagnostic results, ResNet-50 provides outstanding performance for diagnosing ocular fundus diseases on clear fundus images. However, cataracts impact the diagnosis that the risk of misclassification significantly increases in patients with cataracts. 
The effectiveness of the proposed algorithm is illustrated in the performance rebound of the images restored by ArcNet.

\subsection{Model setting analysis}
In this section, various model settings, including image size and HFC, are compared to analyze their impacts on the performance of ArcNet. And the corresponding computational cost and time are also presented.

\begin{figure}[hbt]
\begin{centering}
\includegraphics[width=8.8cm]{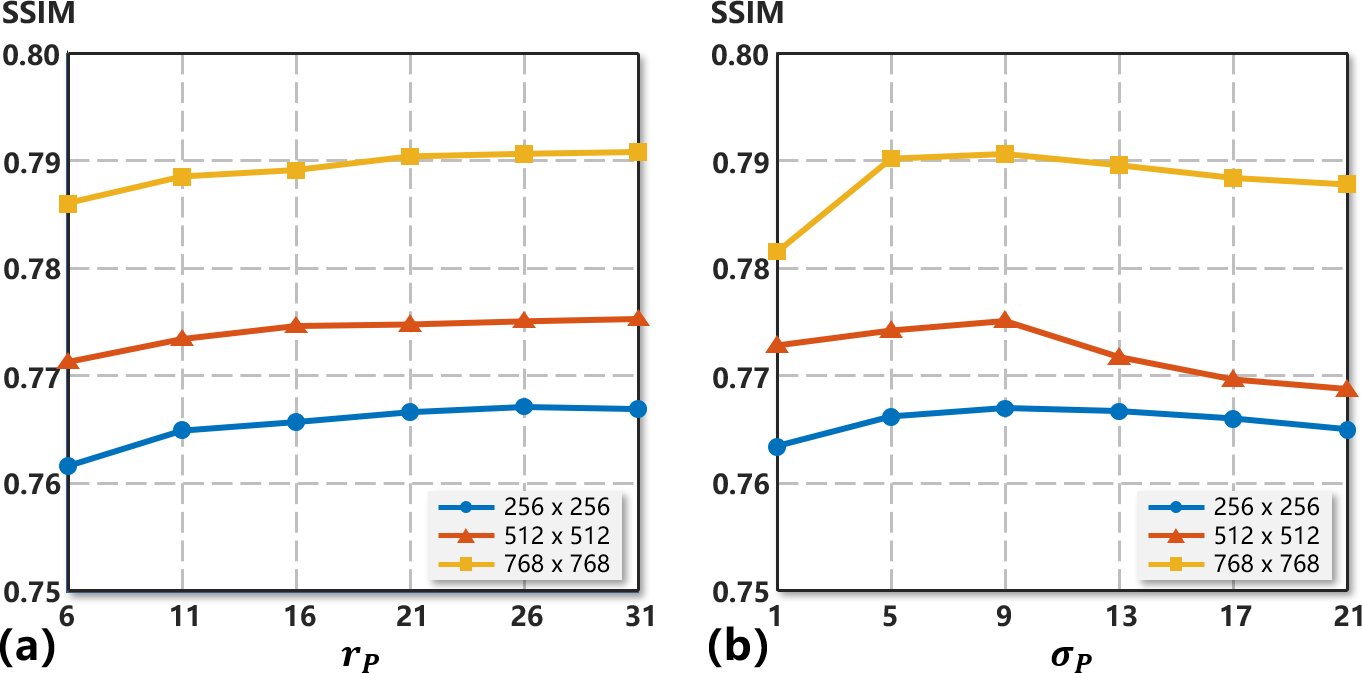}
\par\end{centering}
\caption{Restoration performance with various image sizes and HFC. Three images sizes are provided and the performance of the HFC extracted with different $r_p$ is presented in (a), diverse $\sigma_P$ in (b).} \label{fig:parameter}
\end{figure}

Fig.~\ref{fig:parameter} demonstrates the restoration performance with disparate image sizes and HFCs.
The restoration by ArcNet was implemented in three image sizes, i.e. $256 \times 256$, $512 \times 512$, and $768 \times 768$, while the HFC extracted in Eq.~\ref{eq:lowpass} was adjusted by changing the radius $r_P$ and spatial constant $\sigma_P$ of the low-pass Gaussian filter $g_P$.
Under the identical model setting of HFC in Fig.~\ref{fig:parameter}, the image size of $768 \times 768$ provides the best performance, followed by $512 \times 512$ and $256 \times 256$.
This illuminates that as a result of the increased resolution, the restoration performance grows with the image size.
On the other hand, according to Eq.~\ref{eq:lowpass}, the HFC is obtained through removing the LFC captured by $g_P$.
Therefore, structure details will remain in the LFC and then be removed from the HFC if $r_P$ is not large enough. 
As shown in Fig.~\ref{fig:parameter} (a), the restoration performance improves with the increase of $r_P$, and the performance starts to converge once $r_P$ reaches 26.
Further, consistent with~\cite{getreuer2013survey}, the optimal performance is achieved in Fig.~\ref{fig:parameter} (b) at $\sigma_P$ of 9 cooperating with $r_P$ of 26.

The computational cost, training time, and inference time of ArcNet are summarized in Table~\ref{tab:computer} to demonstrate the overall complexity. The computational cost is quantified by multiply-accumulate operation (GMac), and training and inference time refer to the number of hours (h) for training ArcNet and the number of seconds (s) for restoring an image in the inference phase.
As the image size increases, the computational cost increases exponentially, and the training and inference time grow correspondingly. Combining Fig.~\ref{fig:parameter} and Table~\ref{tab:computer}, it indicates that a larger image size achieves better restoration performance, but the computational complexity accordingly increases.
Thus considering the computational cost, the image size of $256 \times 256$ was employed to conduct the experiments.

\begin{table}[htbp]
\footnotesize
\centering {\caption{Comparison of computational complexity with different image sizes.}
\label{tab:computer} }
\renewcommand{\arraystretch}{1.2}
\begin{tabular}{p{1.4cm}<{\centering} p{0.1cm} p{1.5cm}<{\centering}p{1.2cm}<{\centering}p{1.2cm}<{\centering}}
\hline
Image size && Computational cost (GMac) & Training time (h) & Inference time (s)\\
\hline
$256 \times 256$&& 18.16 & 3.02 & 0.10\\
$512 \times 512$&& 72.81 & 3.22 & 0.32\\
$768 \times 768$&& 163.83  & 3.56 & 0.51\\
\hline
\end{tabular}
\end{table}

\section{Discussion}
According to the experiments, the proposed ArcNet is an efficient annotation-free restoration algorithm for cataractous fundus images. 
Compared to segmentation, HFC is more suitable to preserve retinal structures since it is free from segmentation annotation, and more importantly, the unannotated structure details are favorably preserved by HFC.
The modules in ArcNet allow it to outperforms the state-of-the-art algorithms in the absence of annotations. Notably high-quality fundus images are still necessary to guarantee the performance of ArcNet. In the diagnosis of ocular fundus diseases,  the clinical application value of ArcNet is presented that it promotes the accuracy of automatic diagnosis algorithms.

The annotation of clinical data is a bottleneck for the application of deep learning in many medical scenarios. Developing restoration algorithms for cataractous images requires the annotation from clear images of identical patients, which is extremely troublesome in clinics. Additionally, segmentation is employed in recent algorithms to preserve retinal structures in restoration. Under the prerequisite of segmentation annotations, these algorithms have practical difficulties in clinical deployment. The proposed ArcNet circumvents these rigorous requirements to restore cataractous fundus images without any annotation. Cataract-like images are simulated to train the restoration model, and structure guidance is automatically captured from the HFC of fundus images to preserve the fine retinal structures. The learned model is efficiently adapted to real cataract images by unsupervised domain adaptation.
Thus ArcNet enjoys outstanding clinical practicability.

Although ArcNet circumvents the impractical requirement of annotations for fundus images, collecting high-quality fundus images as the source domain is still significant to guarantee the performance of ArcNet.
In clinical applications, restoring the cataractous images using the source domain synthesized by the clear fundus images obtained under the conditions same as the cataractous ones will help ArcNet achieve the best restoration performance.

\section{CONCLUSION}
Image restoration is a solution to the challenge of retinal examination caused by cataracts in fundus images.
However, the annotation requirement limits restoration algorithms' development and application for cataractous fundus images. To mitigate this limitation, in this paper an annotation-free restoration network, named ArcNet, was proposed to restore cataractous fundus images while preserving the fine retinal structures.  Extensive experiments were conducted to validate the performance and effectiveness of the proposed ArcNet. ArcNet restored cataractous images without annotations and outperformed the state-of-the-art restoration algorithms. Also, in cataracts patients, ArcNet boosted the performance of automatic diagnostic algorithms for ocular fundus diseases.

\bibliographystyle{IEEEtran}
\bibliography{main}

\end{document}